\documentclass[aps,prl,twocolumn,groupedaddress,floatfix,amsmath,showpacs]{revtex4}

\usepackage{graphicx}

\begin{document}
\title{Non-resonant wave front reversal of spin waves used for microwave signal processing}

\author{V. I. Vasyuchka}
\email{vasyuchka@physik.uni-kl.de}
\altaffiliation[\\ Also at ]{National Taras Shevchenko University of
Kiev, Ukraine.}

\affiliation{Fachbereich Physik and Forschungszentrum OPTIMAS, Technische Universit\"{a}t
Kaiserslautern, 67663 Kaiserslautern, Germany}

\author{G. A. Melkov}
\affiliation{Department of Radiophysics, National Taras Shevchenko University of Kiev, 01033 Kiev, Ukraine}

\author{A. N. Slavin}
\affiliation{Department of Physics, Oakland University, Rochester, Michigan 48309, USA}

\author{A. V. Chumak}
\affiliation{Fachbereich Physik and Forschungszentrum OPTIMAS, Technische Universit\"{a}t
Kaiserslautern, 67663 Kaiserslautern, Germany}

\author{V. A. Moiseienko}
\affiliation{Department of Radiophysics, National Taras Shevchenko University of Kiev, 01033 Kiev,
Ukraine}

\author{B. Hillebrands}
\affiliation{Fachbereich Physik and Forschungszentrum OPTIMAS, Technische Universit\"{a}t
Kaiserslautern, 67663 Kaiserslautern, Germany}

\date{\today}

\begin{abstract}
It is demonstrated that non-resonant ($\omega_\mathrm{s}\neq\omega_\mathrm{p}/2$) wave front reversal
(WFR) of spin-wave pulses (carrier frequency $\omega_\mathrm{s}$) caused by pulsed parametric pumping
(carrier frequency $\omega_\mathrm{p}$) can be effectively used for microwave signal processing. When
the frequency band $\Omega_\mathrm{p}$ of signal amplification by pumping is narrower than the spectral
width $\Omega_\mathrm{s}$ of the signal ($\Omega_\mathrm{p}\ll\Omega_\mathrm{s}$), the non-resonant WFR
can be used for the analysis of the signal spectrum. In the opposite case
($\Omega_\mathrm{p}\gg\Omega_\mathrm{s}$) the non-resonant WFR can be used for active (with
amplification) filtering of the input signal.
\end{abstract}


\maketitle

The phenomenon of wave front reversal (WFR), wherein a propagating wave packet of a carrier frequency
$\omega_\mathrm{s}$ can be reversed by external pumping of a carrier frequency
$\omega_\mathrm{p}\simeq\omega_\mathrm{s}$, is well-known in nonlinear optics and acoustics, where it is
caused by a four-wave second-order parametric interaction \cite{Zeldovich85}. For spin waves in
magnetically  ordered substances it is possible to realize WFR in a three-wave parametric process
\cite{Melkov00-PRL-WFR, Serga05-PRL-BulletWFR} with conservation laws:
\begin{equation}\label{conserv}\omega_\mathrm{s}~+~\omega_\mathrm{i}~=~\omega_\mathrm{p},~~~
\textbf{k}_\mathrm{s}~+~\textbf{k}_\mathrm{i}~=~\textbf{k}_\mathrm{p},
\end{equation}
where $\textbf{k}_\mathrm{p}$, $\textbf{k}_\mathrm{s}$,
 are the carrier wave vectors of the pumping
and signal pulses, and $\omega_\mathrm{i}$ and $\textbf{k}_\mathrm{i}$ are the carrier frequency and
wave vector of the ``idle" wave pulse formed as a result of the parametric interaction. When the pumping
wave vector is much smaller than the carrier wave vector of the signal
($\textbf{k}_\mathrm{p}\ll\textbf{k}_\mathrm{s}$) the generated ``idle" pulse  is wave-front reversed,
i.e. $\textbf{k}_\mathrm{i}\simeq-\textbf{k}_\mathrm{s}$.

WFR of microwave spin-wave packets propagating in a thin ferromagnetic film is performed by applying a
pulse of electromagnetic parametric pumping to the film \cite{Melkov00-PRL-WFR, Melkov99-JETP-theory}.
Previously, WFR of spin waves was realized when the the pumping frequency was exactly twice as large as
the carrier frequency of the incident spin-wave packet, i.e. $\omega_\mathrm{p} = 2 \omega_\mathrm{s}$.

Here, we demonstrate that the phenomenon of \emph{non-resonant} WFR, where the input signal frequency is
not exactly equal to half of the pumping frequency ($\omega_\mathrm{s}\neq\omega_\mathrm{p}/2$), opens
interesting new possibilities for signal processing in the microwave frequency range. Spectrum analysis
of microwave pulses using the non-resonant WFR process can be realized when the frequency band
$\Omega_\mathrm{p}$ of the signal amplification by the applied parametric pumping is narrower than the
spectral width $\Omega_\mathrm{s} = 2\pi/\tau_\mathrm{s}$ of the input signal pulse
($\Omega_\mathrm{p}\ll\Omega_\mathrm{s}$), where $\tau_\mathrm{s}$ is the duration of the input signal
pulse. In the opposite limiting case, when $\Omega_\mathrm{p}\gg\Omega_\mathrm{s}$, the non-resonant WFR
can be used for active (with amplification) filtering of input microwave signals.

\begin{figure}[t]
\begin{center}
\scalebox{1}{\includegraphics[width=7.5 cm,clip]{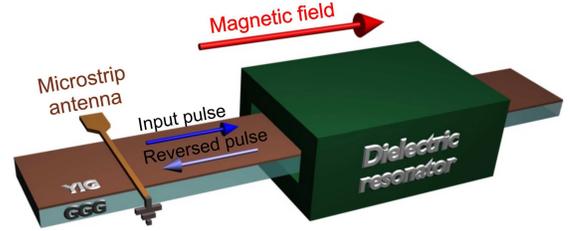}}
\end{center}
\vspace*{-0.4cm}\caption{(Color online) Schematic experimental setup used for the non-resonant wave
front reversal investigation.} \label{Setup}
\end{figure}

A scheme of the setup used in our experiments is shown in Fig.~\ref{Setup}. The input microwave signal
pulse was converted by a microstrip antenna into a packet of dipolar spin waves propagating in a long
(20~mm) and narrow (1.8~mm) 5~$\mu$m thick waveguide made of a ferrimagnetic yttrium iron garnet (YIG)
film epitaxially grown on a gallium gadolinium garnet (GGG) substrate. A bias magnetic field was applied
in the plane of the YIG film along the spin-wave propagation direction. Thus, similar to the earlier
experiment \cite{Melkov00-PRL-WFR}, the backward volume magnetostatic waves (BVMSW) were excited in the
waveguide. The input BVMSW wave packets were excited by microwave pulses of the carrier frequency
$\omega_\mathrm{s}$, power $P_s=0.25$~mW, and duration varying from $\tau_\mathrm{s}=50$~ns to
100~$\mu$s.

The central part of the YIG waveguide was placed inside an open dielectric resonator (ODR), as is shown
in Fig.~\ref{Setup}. The ODR was used for supplying the pumping pulses at a fixed carrier frequency
$\omega_\mathrm{p}$. The pumping magnetic field was oriented parallel to the bias magnetic field,
leading to parallel pumping conditions. Pumping pulses lasting from $\tau_\mathrm{p}=50$~ns to 500~ns
and power of up to 5~W were supplied at time $t_\mathrm{d}$. The parametric interaction of the signal
wave packet with the pumping pulse, which has a carrier frequency close (but not exactly equal) to
$\omega_\mathrm{s}$, created a wave front-reversed wave packet, propagating in the opposite direction.
This wave packet was detected at the input antenna at the time $2t_\mathrm{d}+\tau_\mathrm{p}$ after the
input signal was launched. For further details on the setup see Ref.~\cite{Melkov04-PRB-setup}.

The simplified formalism  previously developed in  Ref.~\cite{Melkov99-JETP-theory} was used for the
theoretical analysis of the observed non-resonant WFR process. In the framework of this formalism,
performed in wave vector space for the case of strong pumping, the ``\emph{k}-th" Fourier component of
the front-reversed spin-wave pulse at the time $t=2t_\mathrm{d}+\tau_\mathrm{p}$, when this component
has a maximum value, can be approximately evaluated as:
\begin{equation}\label{base1}
C_{-k} \approx C_{k0} \cdot \exp{\Big[(V_k h_\mathrm{p} - \Gamma)\tau_\mathrm{p} - \Gamma
(2t_\mathrm{d}+\tau_\mathrm{p}) - \Big(\frac{\Delta \omega_k}{\Omega_\mathrm{p}}\Big)^2\Big]},
\end{equation}
where $C_{k0}$ is the  $k$-th Fourier component of the input (signal) spin-wave pulse, $h_\mathrm{p}$ is
the amplitude of the pumping magnetic field, $V_k$ is the coefficient of parametric coupling between
pumping and spin waves defined in \cite{Melkov99-JETP-theory}, $\Gamma$ is the spin-wave relaxation
parameter, $\Delta \omega_k = \omega_k - \omega_\mathrm{p}/2$ is the detuning between half of the
pumping frequency and the frequency $\omega_k$ of the "\emph{k}-th" Fourier component of the signal, and
\begin{equation}\label{omega-p}
\Omega_\mathrm{p} = \sqrt{2 V_k h_\mathrm{p}/\tau_\mathrm{p}},
\end{equation}
is the bandwidth of parametric amplification of the signal by pumping (see \cite{Melkov99-JETP-theory}
for details).

It is clear from (\ref{base1}) that the amplitude $C_{-k}$ of the front-reversed spin wave increases
with an increasing pumping magnetic field $h_\mathrm{p}$ and is at maximum for $\Delta\omega_k=0$, which
is the case of exact parametric resonance. One can see in (\ref{omega-p}) that the parametric
amplification bandwidth $\Omega_\mathrm{p}$ is determined by the pumping pulse amplitude $h_\mathrm{p}$
and duration $\tau_\mathrm{p}$.

As is pointed out above, the analytic expression (\ref{base1}) is approximate,  and is applicable only
to a qualitative analysis. For a more accurate calculation of the power $P_r$ of the front-reversed
output pulse we performed a numerical summation of the Fourier components (\ref{base1})
($P_\mathrm{r}(t)=|\sum{C_{-k}(t)}|^2$) , taking into account fast phase oscillation and the accurate
dispersion relation for BVMSW \cite{Kalinikos86-dispersions}. The results of this more accurate approach
are shown as theoretical curves in the figures of this paper.

Below, we shall consider two limiting cases  of the non-resonant WFR: the broadband input signal case,
when $\Omega_\mathrm{s}\gg\Omega_\mathrm{p}$, and the narrow-band input signal case, when
$\Omega_\mathrm{s}\ll\Omega_\mathrm{p}$.

\subsection{a) Broadband input signal regime, $\Omega_\mathrm{s}\gg\Omega_\mathrm{p}$}

\begin{figure}[t]
\begin{center}
\scalebox{1}{\includegraphics[width=7.5 cm,clip]{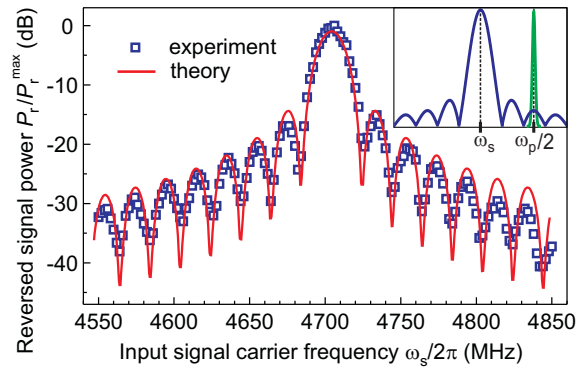}}
\end{center}
\vspace*{-0.4cm}\caption{(Color online) Experimental (symbols) and calculated (line) dependences of the
reversed signal power on the input signal frequency in the regime of broadband input signal
($\Omega_\mathrm{s}/\Omega_\mathrm{p}\approx3.5$). Inset: the spectrum of the input signal (blue line)
and the band of parametric amplification (green line).} \label{Res1}
\end{figure}

The spectral picture of the non-resonant WFR in the case of broadband input signal is shown in the inset
of Fig.~\ref{Res1}. The parametric amplification bandwidth is narrow (it is equal to the delta function
$\delta(\omega-\omega_\mathrm{p}/2)$ in the limiting case), thus only one input signal spectral
component $C_{k0}$ with the frequency equal to half of the pumping frequency $\omega_\mathrm{p}/2)$ can
be amplified by the pumping. The amplitude of the $C_{-k}$ component in the reversed pulse is
proportional to the amplitude of the corresponding component $C_{k0}\mid
_{\omega_k=\omega_\mathrm{p}/2}$ of the input signal. The process looks like ``sampling" of the input
signal spectrum $C_{k0}$ by the spectrally narrow pumping, as it is illustrated in the inset of
Fig.~\ref{Res1}. As a result, the spectrum of the input signal can be directly extracted by the
non-resonant WFR process. The carrier frequency of the output reversed signal in this case is equal to
$\omega_\mathrm{p}/2$.

The results of comparison between the experimentally measured (squares) and theoretically calculated
(solid line) power of the reversed pulse as a function of the carrier frequency of the input (signal)
wave packet are shown in Fig.~\ref{Res1} for the case when the signal duration $\tau_\mathrm{s}$ was
50~ns, pumping duration $\tau_\mathrm{p}$ was 50~ns, and the pumping field multiplied with the coupling
coefficient was $h_\mathrm{p}V_k/2\pi=5$~MHz. The ratio of the characteristic bandwidths was
$\Omega_\mathrm{s}/\Omega_\mathrm{p}\approx3.5$, and, therefore a broadband input signal regime of WFR
was realized. The bias magnetic field was equal to 1020~Oe, the pumping frequency was
$\omega_\mathrm{p}/2\pi=9420$~MHz, the spin-wave relaxation parameter was $\Gamma/2\pi=0.42$~MHz, and
the delay time between the signal and pumping was $t_\mathrm{d}=65$~ns.

It is clear from Fig.~\ref{Res1} that theory is in excellent agreement with experiment.

We would like to mention, that an attempt to use non-resonant three-wave parametric interaction of spin
waves for microwave spectrum analysis was undertaken in a recent paper
\cite{Schaefer08-APL-Restoration}. However, the interaction process used in
\cite{Schaefer08-APL-Restoration, PRL-Serga-07} involves several different spin-wave groups and, as a
result, the information about the duration, and shape of the input signal pulse was lost, and the
amplitude of the obtained ``spectrum" of the input pulse was considerably modified. This is not the case
in the results reported here.

\subsection{b) Narrow-band input signal regime, $\Omega_\mathrm{s}\ll\Omega_\mathrm{p}$}

\begin{figure}[t]
\begin{center}
\scalebox{1}{\includegraphics[width=7.5 cm,clip]{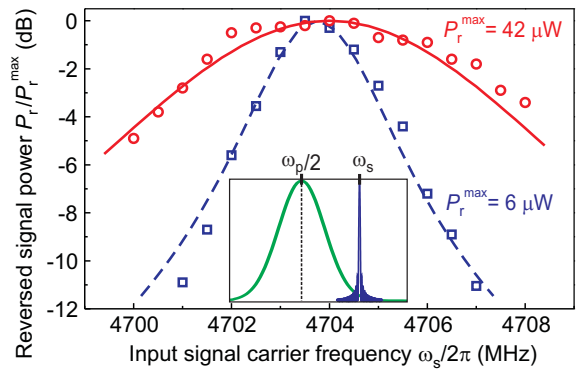}}
\end{center}
\vspace*{-0.4cm}\caption{(Color online) Experimental (symbols) and calculated (lines) dependences of the
reversed signal power on the input signal frequency in the regime of narrow-band input signal. Circles
and solid line correspond to the pumping power $P_\mathrm{p}\approx8$~mW and pumping duration
$\tau_\mathrm{p}=500$~ns ($\Omega_\mathrm{s}/\Omega_\mathrm{p}\approx0.01$). Squares and dashed line
correspond to the pumping power $P_\mathrm{p}\approx200$~mW and pumping duration
$\tau_\mathrm{p}=100$~ns ($\Omega_\mathrm{s}/\Omega_\mathrm{p}\approx0.0025$). Both curves were
normalized by corresponding maximum powers. Inset: the band of parametric amplification (green line) and
the spectrum of input signal (blue line).} \label{Res2}
\end{figure}

The opposite limiting case of a narrow-band input signal regime of WFR is realized when the input signal
pulse is sufficiently long (or/and when the pumping pulse is sufficiently short and strong), i.e when
the condition $\Omega_\mathrm{s}\ll\Omega_\mathrm{p}$ is fulfilled. It is clear that in this case the
spectral width of the signal is much smaller than the band of frequencies amplified by pumping (see
inset in Fig.~\ref{Res2}). The amplitude of the reversed pulse is determined by the amplitude of the
main harmonic of the input signal, by the detuning ($\omega_\mathrm{s}-\omega_\mathrm{p}/2$), and by the
parametric amplification bandwidth $\Omega_\mathrm{p}$.

The measured (symbols) and theoretically calculated (lines) power of the reversed pulse as a function of
the carrier frequency of the input signal obtained for two different pumping pulses are compared in
Fig.~\ref{Res2}. In both cases the theoretical and experimental curves were normalized by the
corresponding maxima of experimentally measured power of the reversed pulse. The duration of the input
signal pulse was  $\tau_\mathrm{s}=100~\mu$s. The bias magnetic field was 950~Oe and the pumping carrier
frequency was $\omega_\mathrm{p}/2\pi=9408$~MHz.

We performed experiments for two pumping amplitudes and durations. In the first case (circles and solid
line in Fig.~\ref{Res2}) the duration of the pumping pulse $\tau_\mathrm{p}$ was 500~ns and the pumping
field multiplied with the coupling coefficient was $h_\mathrm{p}V_k/2\pi=1$~MHz (corresponding to the
pumping power of $P_\mathrm{p}\approx8$~mW). The characteristic ratio
$\Omega_\mathrm{s}/\Omega_\mathrm{p}$ was equal to 0.01 for these conditions. In the second case
(squares and dashed lines in Fig.~\ref{Res2}) the parameters were: $\tau_\mathrm{p}=100$~ns and
$h_\mathrm{p}V_k/2\pi=5$~MHz (corresponding to $P_\mathrm{p}\approx200$~mW), and the characteristic
ratio $\Omega_\mathrm{s}/\Omega_\mathrm{p}$ was equal to 0.0025. Thus, a narrow-band input signal regime
of the WFR process was realized for both pumping pulses.

The curves presented in Fig.~\ref{Res2} can be interpreted as the amplitude-frequency characteristics of
an active parametric pass-band microwave filter i.e. tunable by a pumping filter with amplification of
the filtered signal. The bandwidth and the gain factor of the filter can be tuned by varying the
duration and the power of the applied pumping pulse. Indeed, for the first pumping pulse (circles in
Fig.~\ref{Res2}) the pass-band of the active filter at the -3 dB level is equal to 6.4~MHz and for the
second pumping pulse (squares in the figure) this pass-band is 2.4~MHz. Note that the gain coefficient
of the active filter is substantially higher for the first pumping pulse than for the second one, but
the curves shown in Fig.~\ref{Res2} were normalized. It should also be mentioned that in order to
achieve good quantitative agreement between the theory and experiment shown in Fig.~\ref{Res2} it was
necessary to perform the full summation of all the Fourier components of the interacting pulses and
release the condition of strong pumping. The approximate analytic formula Eq.~\ref{omega-p} describes
the active filtering process only qualitatively. Nevertheless, it correctly predicts the decrease of the
filter pass-band with the increase of the power and decrease of duration of the pumping pulse.

In conclusion, we explicitly demonstrated that non-resonant  wave front reversal for dipolar spin waves
can be effectively used for spectral analysis and active filtering of pulsed microwave signals. We
believe that signal processing devices based on parametric interaction of spin waves in magnetic films
will find important applications in microwave technology.

The authors would like to thank Dr. A. A. Serga for helpful discussions and Dr. S. Trudel for a critical
reading. Financial support from the DFG within the SFB/TRR 49, Ukrainian Fund for Fundamental Research
(25.2/009), National Science Foundation of the USA (grant No. ECCS 0653901), U.S. Army TARDEC, RDECOM
(contract N0.W56HZW- 09-P-L564), and the U.S. Army Research Office (grant No. W911NF-04-1-0247) is
gratefully acknowledged.

\end{document}